# Experimental Study of Ethylene Evaporites Under Titan Conditions

*Ellen C. Czaplinski[a*], Woodrow A. Gilbertson[b], Kendra K. Farnsworth[a], and Vincent F. Chevrier[a]*

[a] Arkansas Center for Space and Planetary Sciences, University of Arkansas, Fayetteville, AR 72701, USA
[b] Department of Physics, University of Arkansas, Fayetteville, AR 72701, USA

We wish to confirm that there are no known conflicts of interest associated with this publication and there has been no significant financial support for this work that could have influenced its outcome.

We confirm that the manuscript has been read and approved by all named authors and that there are no other persons who satisfied the criteria for authorship but are not listed. We further confirm that the order of authors listed in the manuscript has been approved by all of us.

We understand that the Corresponding Author is the sole contact for the Editorial process (including Editorial Manager and direct communications with the office). They are responsible for communicating with the other authors about progress, submissions of revisions, and final approval of proofs. We confirm that we have provided a current, correct email address which is accessible by the Corresponding Author and which has been configured to accept email from: ecczapli@email.uark.edu

Signed by all authors as follows:

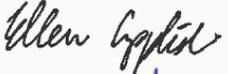

Ellen Czaplinski

Vincent Chevrier

Kendra Farnsworth

Woodrow Gilbertson

ABSTRACT

Titan has an abundance of lakes and seas, as confirmed by Cassini. Major components of these liquid bodies include methane ($CH_4$) and ethane ($C_2H_6$), however, evidence indicates that minor components such as ethylene ($C_2H_4$) may also exist in the lakes. As the lake levels drop, 5-$\mu$m-bright deposits, resembling evaporite deposits on Earth, are left behind. Here, we provide saturation values, evaporation rates, and constraints on ethylene evaporite formation by using a Titan simulation chamber capable of reproducing Titan surface conditions (89 - 94 K, 1.5 bar $N_2$). Experimental samples were analyzed using FTIR (Fourier-Transform Infrared) spectroscopy, mass, and temperature readings. Ethylene evaporites form more quickly in a methane solvent than an ethane solvent, or a mixture of methane/ethane. We measured an average evaporation rate of $(2.8 \pm 0.3) \times 10^{-4}$ kgm$^{-2}$s$^{-1}$ for methane, and an average upper limit evaporation rate of $< 5.5 \times 10^{-6}$ kgm$^{-2}$s$^{-1}$ for ethane. Additionally, we observed red shifts in ethylene absorption bands at 1.630 $\mu$m and 2.121 $\mu$m, and the persistence of a methane band at 1.666 $\mu$m.

## INTRODUCTION

Titan's complex lakes of liquid methane/ethane have intrigued scientists for over a decade. In addition to the lakes and seas we observe today, Titan's past details a history of intricate evaporation processes. The identification of probable evaporite deposits at Titan's north pole, south of Ligeia Mare[1,2] included several "dry" lakebeds showing a unique signature that was bright in the 5-μm window of the Visual Infrared Mapping Spectrometer (VIMS).

These 5-μm-bright regions have been extensively studied and classified as non-water ice materials.[3,4] The 5-μm-bright deposits discovered at Ontario Lacus, described as "bathtub rings" of low water ice condensates, may have been deposited in the past when lake levels were higher.[5,6] Although the exact composition of these 5-μm-bright regions is unknown, previous studies indicate they may be evaporitic in origin.[2,4,7–9] Evaporites form when dissolved solids precipitate out of a saturated solution as that liquid solvent evaporates. Evaporation of the solvent causes the solute to deposit as an evaporite either onto the surface (if all liquid has evaporated), or at the bottom of the saturated liquid (if not all liquid has evaporated).[9] We therefore sought to study Titan-relevant evaporite materials in the laboratory to better constrain the processes that may be occurring in and around Titan's lakes.

Recent evaporite studies have included models and theoretical work,[7,8] and some groups are also experimentally working to constrain potential solvents and solutes that may be active in evaporite formation.[10–15] These previous studies have focused on a number of potential evaporite compounds including: $C_6H_6/C_2H_6$ and $C_2H_2/NH_3$, and $C_2H_2/C_2H_4$. Roe et al.[16] observed a significant increase of ethylene in Titan's polar stratosphere (late southern spring time), suggesting that ethylene may be a common compound in Titan's polar regions, and a possible constituent dissolved in the polar lakes.[17]

However, the high solubility values of ethylene in methane/ethane indicate that the lakes contain less ethylene than its saturation value.[17]

Here, we focus on ethylene ($C_2H_4$) as a potential evaporite, dissolved in liquid methane, ethane, and methane/ethane. The objective of this paper is to provide laboratory context for ethylene evaporite studies on Titan to aid in the understanding of evaporite composition, which is currently unknown. We note that studying a single evaporite compound (e.g. ethylene) is a simplification of what we presume to occur on Titan, however it is important to isolate evaporation rates and detection requirements of simple mixtures before studying more complex scenarios.

EXPERIMENTAL METHODS

**Titan Simulation Chamber.** Experiments were performed using a facility at the University of Arkansas that was designed to simulate surface conditions on Titan.[18] This facility consists of the larger host chamber (Andromeda) with a smaller subsection (the Titan module) where the Temperature Control Box (TCB) sits (Figure 1). Titan-relevant temperatures (89 – 94 K) were produced by liquid nitrogen flow through coils located both inside and outside of the TCB, and inside the condenser. A 1.5 bar atmosphere was maintained via pressurized nitrogen. Once adequate temperatures were reached, the sample was introduced to the condenser where it condensed, then exited the bottom of the condenser through a filter and remained in a sample dish (15 cm diameter x 2 cm deep). We collected Fourier-Transform Infrared (FTIR) spectra of the sample using a Nicolet® 6700 FTIR spectrometer (equipped with a TEC InGaAs® 2.6 μm detector and $CaF_2$ beam-splitter) operating from 1 to 2.6 μm with a spectral sampling of 4 cm$^{-1}$, spectral resolution of 0.01 cm$^{-1}$, and connected to a fiber optic probe located above the sample dish. The bottom of the dish is covered by Spectralon® (from LabSphere), a fluoropolymer that has its highest diffuse (Lambertian) performance in the IR portion of the spectrum, which serves as the background for all FTIR measurements. Sample mass was continuously measured by a Sartorius® GE812 precision balance (precision of 0.01 g). Temperatures of the sample mixture and TCB were constantly monitored via eight K-type (+/- 2.2°C) thermocouples placed throughout (two in the condenser, two on either side of the condenser, one near the FTIR probe, one in the filter, one just above the sample dish, and one directly touching the mixture in the sample dish) and connected to a USB data acquisition module (see Fig. 1 for placement).

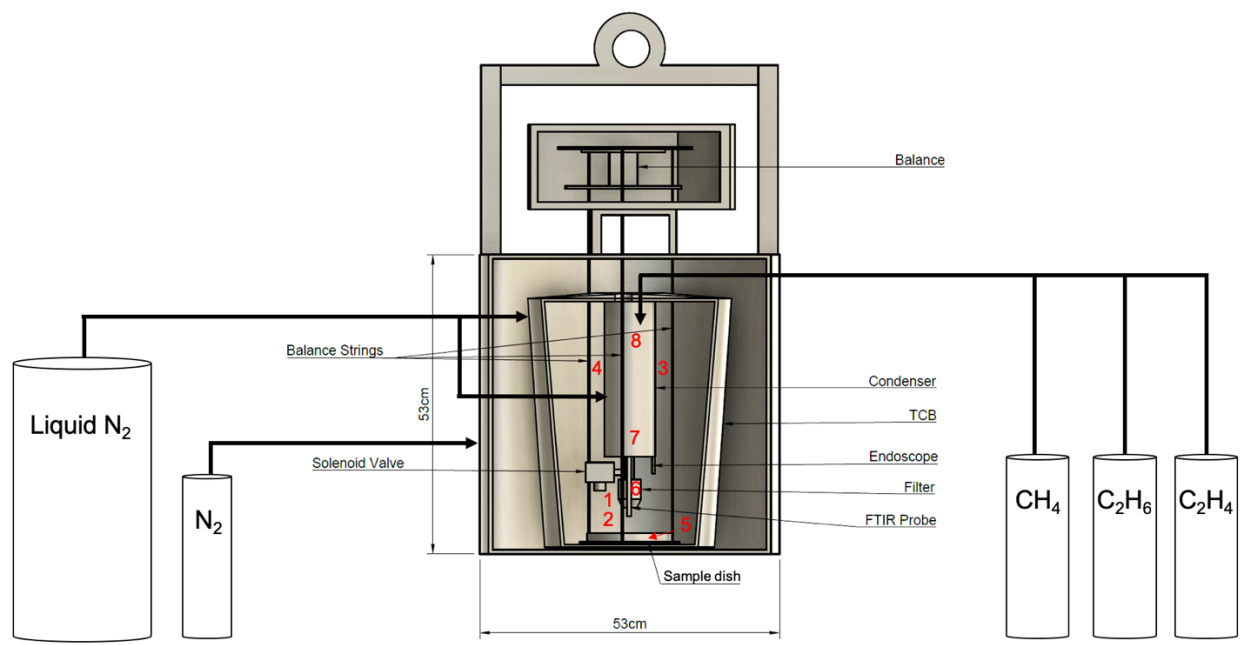

**Figure 1.** Schematic view of the Titan simulation chamber showing a cross-section of the Titan module, which includes the balance and temperature control box (TCB) (to scale). Gas cylinders and feedthroughs are for reference and are not to scale. Red numbers correspond to the locations of different thermocouples. 1 is attached to the FTIR probe (~3 cm above sample), 2 is the atmospheric temperature above the sample (~2 cm), 3 and 4 are the upper atmosphere temperature (~15-20 cm above the sample), 5 is the sample temperature, 6 is the filter temperature, and 7 and 8 are the condenser temperatures. (Modified from Farnsworth et al. (submitted)).

**Experimental Protocol.** To perform evaporite experiments, we began by purging the Titan chamber with nitrogen (Airgas® industrial grade, >99.998%) for 10 minutes[17] to remove any contaminants from previous experiments. Next, the condenser was independently purged with nitrogen by opening the solenoid and exhaust valves for 10 minutes[17] to remove any contaminants. We also maintained a pressure of 1.5 bar in the chamber with nitrogen, monitored by a Matheson 63-3161 pressure gauge (0 – 60 psi). After both purges, we started the flow of liquid nitrogen, which cools both the TCB and condenser. We then collected a background FTIR spectrum of the Spectralon® on the sample dish using Omnic™ FTIR software. Next, we started recording temperature and mass data using LabVIEW software. Experimental time (in minutes), which will be referenced herein, is synchronized with the initiation of the LabVIEW software.

During evaporite experiments, ethylene must dissolve in the solvent mixture before it is introduced to the sample dish. To achieve this, we first injected the gaseous solute (Airgas® ultra high purity, 99.9% ethylene) into the condenser, where it condenses to liquid phase below ~104 K. We then introduced gaseous methane (Airgas® chemically pure grade, 99.5%), ethane (Airgas® chemically pure grade, 99.5%), or both into the condenser for approximately 10 seconds, to undergo phase transition below ~113 K. For reference, at 1.5 bar, methane and ethane are in the liquid phase from ~123 K to 90 K and ~ 184 K to 90 K, respectively. After dissolving ethylene throughout a 10 minute equilibration period,[17] a solenoid valve was opened, allowing the mixture to exit the bottom of the condenser through a glass-fritted filter (40-60 μm pore size). This filtering process ensures that no solid particles >40 μm in diameter travel through to the sample dish, and was monitored with an endoscope camera. The solenoid valve was closed after all the liquid exited the condenser. After the mixture was transferred to the sample dish, FTIR spectral measurements were acquired every ~10 minutes with a resolution of 4 cm$^{-1}$ and an average of 450 scans. Titan surface temperatures (~89 K – 94 K) were sustained after the mixture was poured onto the Spectralon® in the sample dish to simulate Titan's surface temperature and pressure.

It is important to note that our experimental setup only allows approximate abundances of gases to be added during the experiment, and it is not until after the experiment is finished and the spectra are analyzed by our spectral unmixing model that we can determine the exact percentages of each compound in the experiment (See Spectral Unmixing Model Section). We calculated band depths using the Omnic software, which outputs relative reflectance values. Error from these band depth measurements is from the average noise of the spectrometer, which was calculated by taking the standard deviation of the reflectance values from 1.45 – 1.55 μm, as this is a relatively "flat" portion of our experimental spectra with no absorption bands. These standard deviation values were inserted as error bars to each respective sample point on the band depth graphs.

Evaporite detection required that the solvent (ethane/methane) evaporated gradually, leaving behind some residual solute (ethylene). For the methane experiments, we maintained Titan temperatures in the TCB because methane readily evaporates at Titan conditions.[19,20] However, ethane evaporation on the timescales of our experiments is negligible,[20] therefore to induce ethane evaporation, the temperature of the sample was increased. For ethane/ethylene mixtures, we maintained Titan temperatures (~94 K) for one experiment, and slowly warmed the sample to ~139 K for a second experiment. Although ethylene is in liquid phase at the warmer temperatures during the forced evaporation experiment, we note that this could be a local mechanism to enrich a lake in ethylene, while further cooling of the lake may allow ethylene to precipitate.

**Spectral Unmixing Model.** Various spectra of the pure components (CH₄, C₂H₆, C₂H₄) were recorded before use in these experiments (Fig. 2). The pure spectra recorded included both liquid and solid phases of ethylene and ethane, and the liquid phase of methane. Any compound mixture should have a spectrum that is some combination of the spectra of its components. We wrote a Python code to decompose a given spectrum into a best fit of linear combinations of pure spectra added together. This technique provided the weighted composition of the mixture throughout the evaporation process. For compound spectra ($C$) with $i$ components, the Python code breaks down $C$ into pure spectra ($P$) as follows:

$$C = A + B\lambda + \sum_i x_i P_i$$

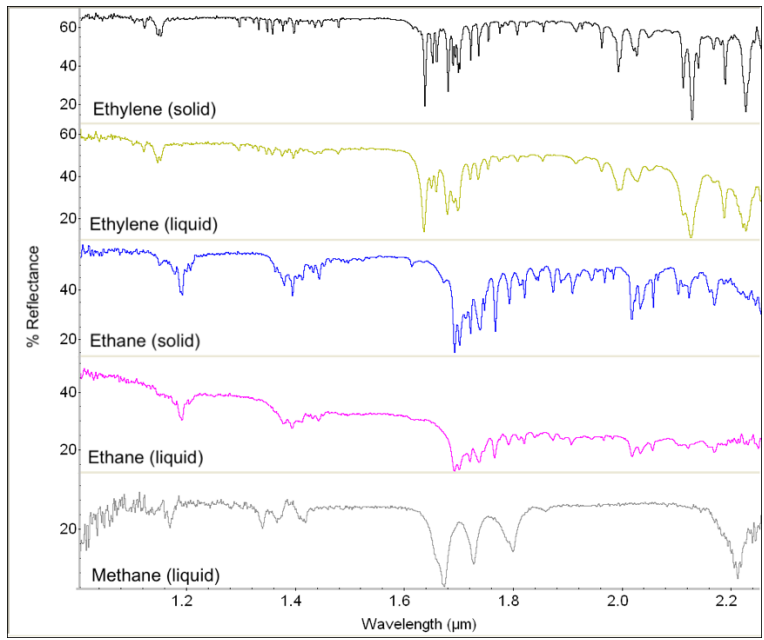

**Figure 2.** Pure spectra from each species used in this study. (Black) solid ethylene, (gold) liquid ethylene, (blue) solid ethane, (magenta) liquid ethane, (gray) liquid methane.

The $A$ and $B\lambda$ terms allowed for the correct of a constant offset and baseline slope, respectively. The spectra tended to increase in reflectance throughout the duration of an experiment, which lead to a vertical offset. Due to instrumental error there was also an occasional slope to the baseline of the reflectance, which was corrected with a simple wavelength ($\lambda$) dependent linear term. The mole fraction of a pure spectrum ($x_i$) was normalized so the sum added to 100%. This process is similar to previous studies,[17] but is an original process for these experiments. The reflectivity of a component was assumed to be proportional to the number of moles of that component, thus the mole fraction of each pure spectrum can be calculated from the compound spectrum. The component percentages listed in the Results section are all the mole fractions given by this analysis. Also listed is the goodness of fit (Table S1), which is defined by total chi-squared value divided by the degrees of freedom. All of our values are close to 1, which shows that the spectral unmixing model is doing an acceptable job. Values significantly lower than 1 would be a sign of over-fitting the data, and likewise values significantly larger than 1 would be a sign of under-fitting the data. The largest final value for goodness of fit (1.37) shows that our model is, at worst, significant with a critical value of 2.5%. Other goodness of fit values all show significance with critical values of 1%.

We assume that this is an accurate method, as the pure spectra were recorded in both liquid and solid phases stable at Titan temperature and pressure in our chamber. This allowed us to model compound mixtures that are both liquid-liquid, as well as solid-liquid to capture the formation of evaporites. The only error introduced would come from the spectrometer. This error is largely accounted for in the $B\lambda$ term, and utilized in the chi-squared minimization used to find the best fit.

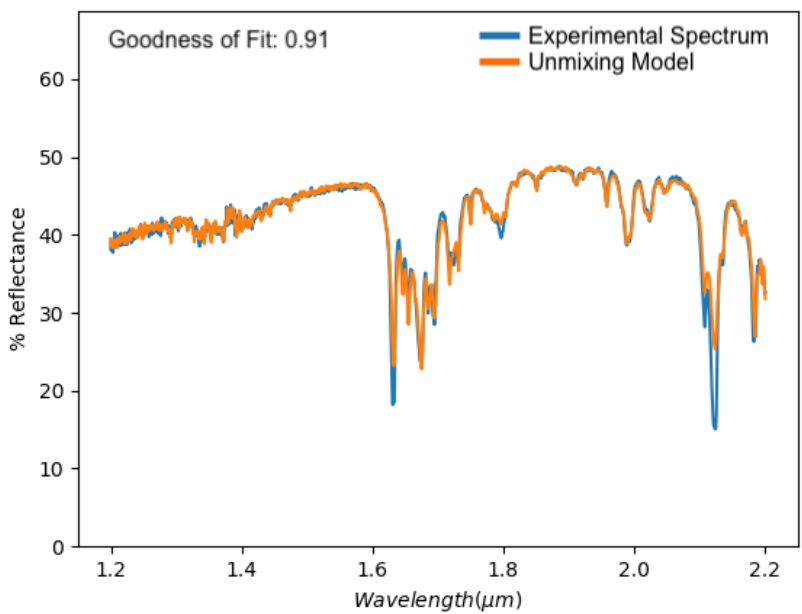

**Figure 3.** Example of a spectrum from the unmixing model (orange) compared to an experimental spectrum (blue) from a methane/ethylene experiment. This was during the methane evaporation phase with methane and ethylene components at 57% and 43%, respectively. Our model accurately fits highly mixed spectra, as shown.

RESULTS

Here, we present NIR reflectance spectra for three types of example experiments: methane/ethylene, ethane/ethylene, and methane/ethane/ethylene. Each experiment has been repeated several times. Additional spectral, mass, and temperature data from repeated experiments can be found in supporting information. Table 1 shows fundamental vibrational frequencies and band assignments for these compounds of interest, as they will be referenced to throughout the paper. Because the edges of our wavelength range show significant noise, we focus on the range from ~1.3 – 2.3 μm, for clarity.

**Table 1.** Spectral band assignments for the compounds used in this study.

| Compound | Assignment | Sample Condition | Freq (cm⁻¹) | Freq (μm) | This Study (μm) |
|---|---|---|---|---|---|
| $CH_4$ | $2\nu_3$ | $CH_4$ in liquid Ar[#] | 5991 | 1.669170422 | 1.669 |
| | $\nu_1/\nu_3 + \nu_2 + \nu_4$ | $CH_4$ in liquid Ar[#] | 5805 | 1.722652885 | 1.723 |
| | | $CH_4$ liquid[#] | 5801 | 1.723840717 | |
| | $\nu_1 + 2\nu_4$ | $CH_4$ in liquid Ar[#] | 5573 | 1.794365692 | 1.796 |
| | | $CH_4$ liquid[#] | 5564 | 1.797268152 | |
| $C_2H_6$ | $\nu_7 + \nu_{10}$ | room temp, 30 mbar[^] | 5948.338 | 1.681141858 | 1.688 |
| | $2\nu_1$ | room temp, 30 mbar[^] | 5901.3 | 1.694541881 | 1.697 |
| | | | | | 2.014 |
| | | | | | 2.03 |
| $C_2H_4$ | $\nu_5 + \nu_9$ | $C_2H_4$ in liquid Ar[+] | 6142.5 | 1.628001628 | 1.632 (liq) |
| | | $C_2H_4$ at 110 K[+] | 6126.6 | 1.632226684 | 1.634 (sol) |
| | $\nu_9 + \nu_2$ | $C_2H_4$ in liquid Ar[+] | 4723.2 | 2.117208672 | 2.123 (liq) |
| | | $C_2H_4$ at 110 K[+] | 4710.6 | 2.122871821 | 2.125 (sol) |

[#]From Blunt et al.[21]
[^]From Hepp and Herman[22]
[+]From Brock et al.[23]

**Methane and Ethylene.** For this experiment, we added 76.3% methane and 23.7% ethylene to the condenser. Offset reflectance FTIR spectra from these experiments are shown in Figure 4 (left). The red spectrum in Figure 4 represents the first spectrum of the experiment (t=81 min), immediately after the mixture was poured onto the sample dish from the condenser. The black spectrum shows an intermediate spectrum (t=136 min). The blue spectrum in Figure 4 represents the final spectrum of the experiment (t=230 min), which ended with 2.0% methane and 98.0% ethylene. Throughout the experiment, the baseline reflectance increased, starting around 40% and ending at approximately 80%.

Mass data for the methane/ethylene experiment shows that the sample was poured in the dish at ~70 minutes, reaching a maximum mass of ~11.3 grams before methane began evaporating after ~90 minutes (Fig. 4, right). Herein, we note that all reported mass data has been corrected for a slight mass drift inherent to the balance at these extremely low temperatures (described in detail in Supporting Information). We report a methane evaporation rate of $(2.6 \pm 0.4) \times 10^{-4}$ $kgm^{-2}s^{-1}$ (Table 2), which is consistent with previously reported experimental values on pure methane evaporation.[19,24]

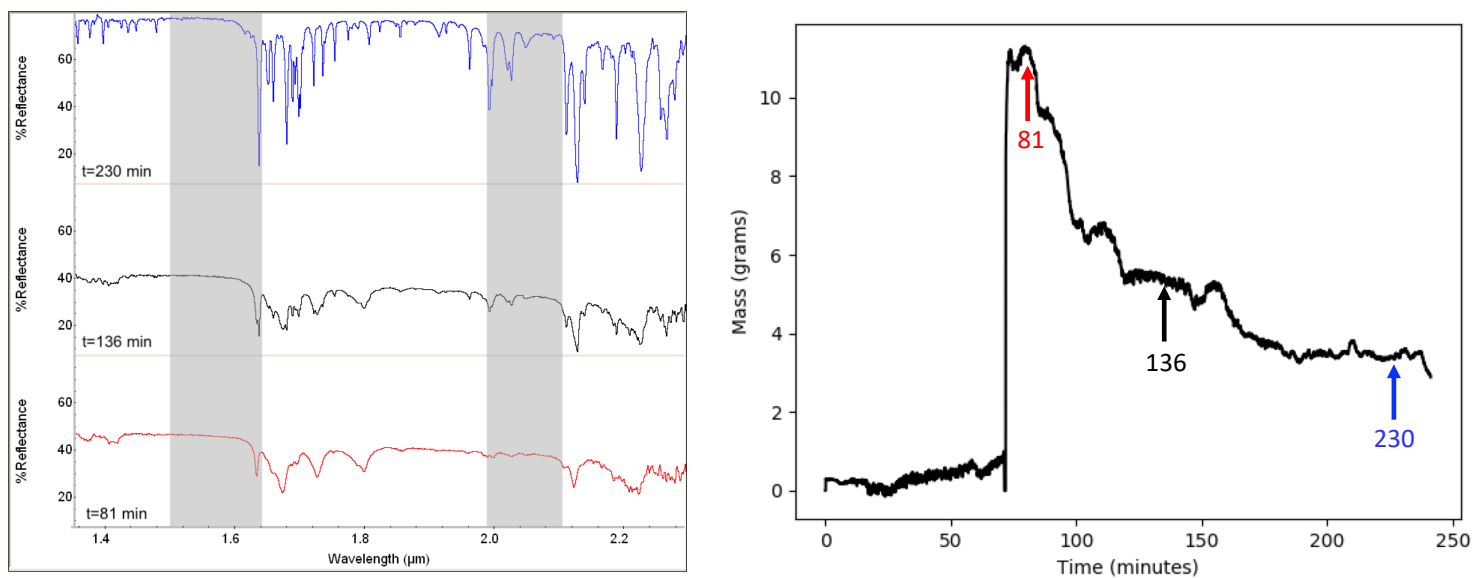

**Figure 4.** Methane/ethylene spectra, offset for clarity (left). Initial spectrum is shown in red, intermediate in black, and final in blue. Gray vertical bars indicate VIMS windows. Notice the increase in reflectance throughout the experiment. Mass data from the methane/ethylene experiment (right). Mass was approximately zero before the mixture was poured in the sample dish (~70 min), and reached a maximum mass of ~11.3 grams. Methane evaporated throughout the experiment, as seen by the almost immediate decrease in mass after ~90 minutes. Colored arrows correspond with the time each spectrum was taken.

Band depth calculations were performed on the major absorption bands of methane (1.669 μm, 1.723 μm, 1.796 μm) and ethylene (1.630 μm, 2.121 μm). As experiment time increased, the methane bands decreased in depth, while the ethylene bands increased in depth. The 1.669 μm methane band decreased in relative reflectance from 9.9% to 0.0%, the 1.723 μm methane band decreased from 11.0% to 0.3%, and the 1.796 μm methane band decreased from 8.3% to 0.0% (Fig. 5). Ethylene bands at 1.630 μm and 2.121 μm increased in relative reflectance from 12.2% to 57.6%, and 9.2% to 49.7%, respectively (Fig. 5). Notice how this increase coincides with the lowest relative reflectance value of methane before reaching zero. Additionally, the decrease in methane band depths throughout the experiment is relatively linear unlike the sharp, fourfold increase observed in the ethylene band depths at ~150 minutes (Fig. 5).

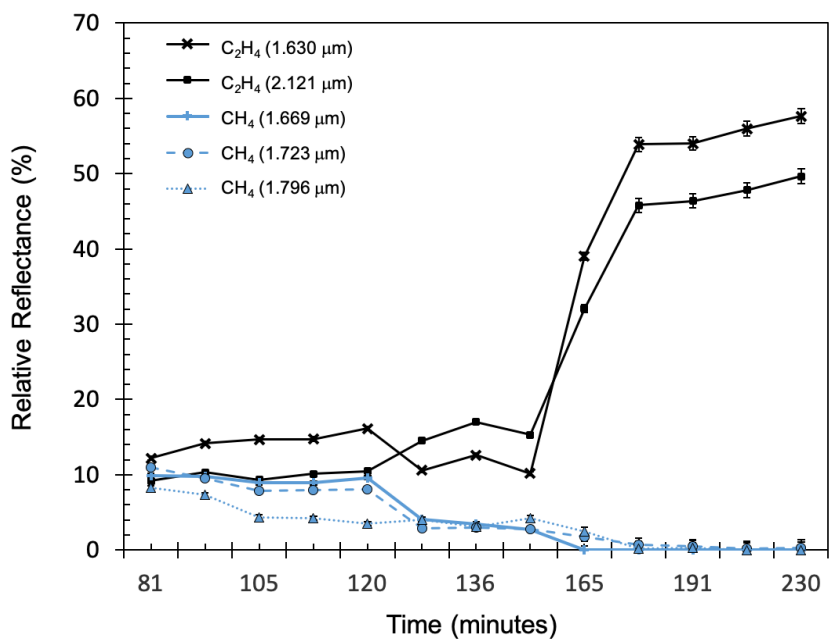

**Figure 5.** Methane band depths (blue) from the methane/ethylene experiment show a decrease in characteristic methane bands (1.669 μm, 1.723 μm, 1.796 μm), while ethylene band depths (black) show an increase in characteristic ethylene bands (1.630 μm, 2.121 μm). The sudden increase of ethylene band depths at ~150 minutes most likely indicates saturation. Error bars are the same size or smaller than point markers.

We observed red shifts of the 1.630 μm and 2.121 μm ethylene bands by 0.003 μm and 0.004 μm, respectively (Fig. 6A,B). Given that the spectral resolution of the spectrometer is 0.01cm$^{-1}$ and the 0.003 μm and 0.004 μm red shifts correspond to shifts of 11.271 cm$^{-1}$ and 8.875 cm$^{-1}$, respectively, these red shifts are well within the resolvable range of the spectrometer. The last spectral measurement taken is shown by the blue spectrum (Fig. 4, left), and nearly matches the pure ethylene spectrum (Fig. 2). It is not an exact match because a band at 1.666 µm is still present in the final spectrum of the experiment (Fig. 6C). This 1.666 µm band originates from the first overtone region of the pure methane spectrum, but does not disappear throughout the duration of the experiment like all other methane bands.

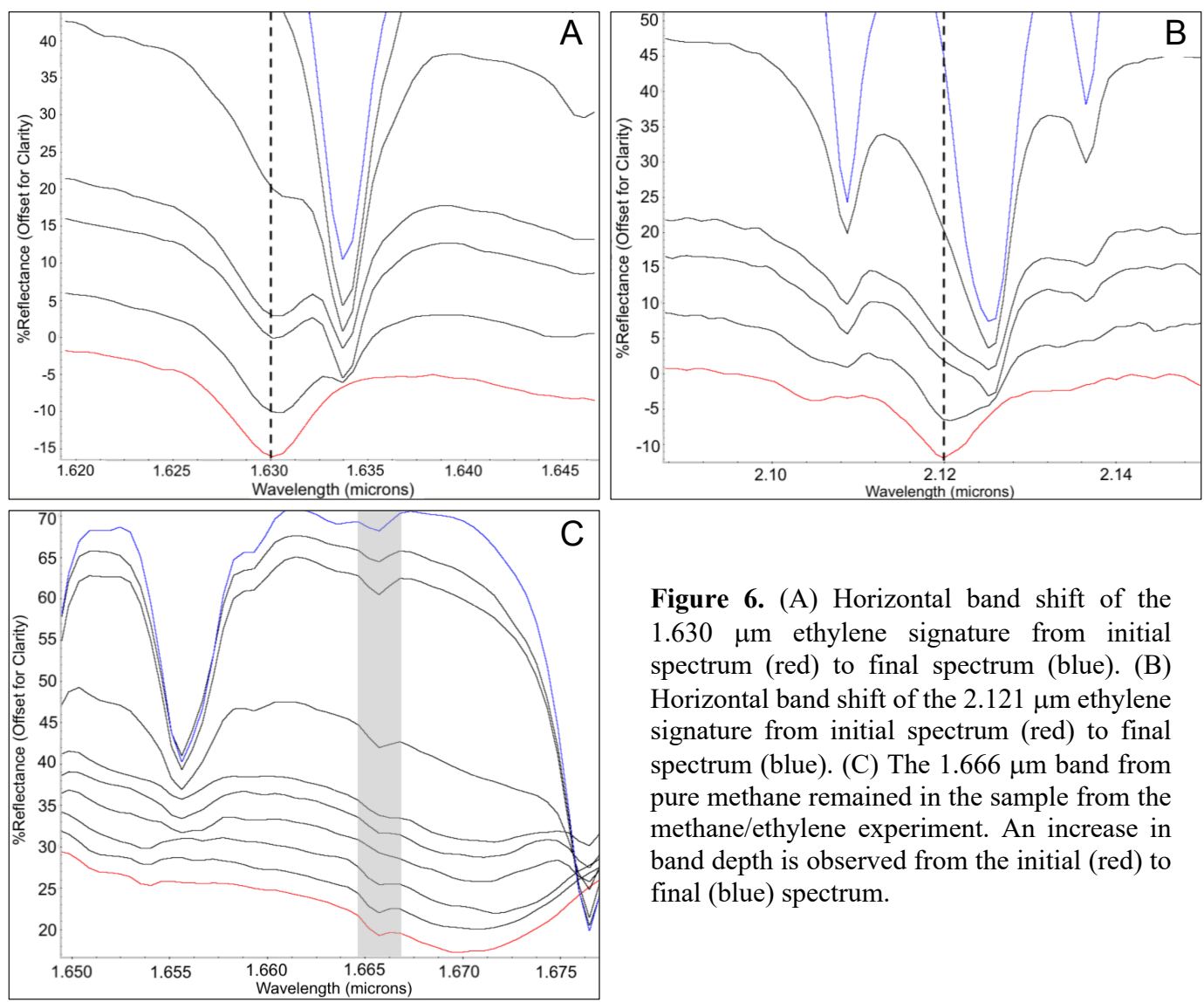

**Figure 6.** (A) Horizontal band shift of the 1.630 µm ethylene signature from initial spectrum (red) to final spectrum (blue). (B) Horizontal band shift of the 2.121 µm ethylene signature from initial spectrum (red) to final spectrum (blue). (C) The 1.666 µm band from pure methane remained in the sample from the methane/ethylene experiment. An increase in band depth is observed from the initial (red) to final (blue) spectrum.

**Ethane and Ethylene**

**90 K Experiment.** This experiment consisted of 87.9% ethane and 12.1% ethylene, initially. Figure 7 (left) shows the offset FTIR spectra. The first spectrum was taken immediately after the mixture was poured onto the sample dish (t=95 min), and is indicated by the red spectrum in Figure 7. An intermediate spectrum was taken at t=132 min. The final spectrum for this experiment is highlighted in blue in Figure 7 (t=187 min) and contained 76.0% ethane, and 24.0% ethylene. The spectrum in Figure 7 shows an approximately constant baseline reflectance around 35% - 40%, and minor variations in band depth.

Figure 7 (right) shows the mass data for this experiment. This experiment included two pours due to nonideal chamber conditions after the first pour. The analyzed ethane/ethylene sample was poured onto the petri dish ~85 minutes after the start of the experiment. After the pour, the mass remained constant (~20 g). We report an upper limit ethane evaporation rate of < 7.0 x $10^{-6}$ kgm$^{-2}$s$^{-1}$ (Table 2), which is consistent with the upper limit previously reported by Luspay-Kuti et al..[20]

No red shifts of any bands are present in the spectra when comparing to the pure endmembers of ethane and ethylene (Fig. 2). Band depth calculations (Fig. 8) were performed on ethane bands (1.688 μm, 1.697 μm, 2.014 μm, and 2.030 μm) and ethylene bands (1.630 μm, 2.121 μm). As time increased, ethylene band depths consistently increased, while ethane band depths remained constant throughout the experiment. The average relative reflectance values of the 1.688 μm, 1.697 μm, 2.014 μm, and 2.030 μm ethane bands were 4.0%, 2.3%, 3.9%, and 2.3%, respectively (Fig. 8). The 1.630 μm ethylene band depths increased in relative reflectance from 1.61% to 6.59%, whereas the 2.121 μm ethylene band depths increased from 1.7% to 3.0% (Fig. 8).

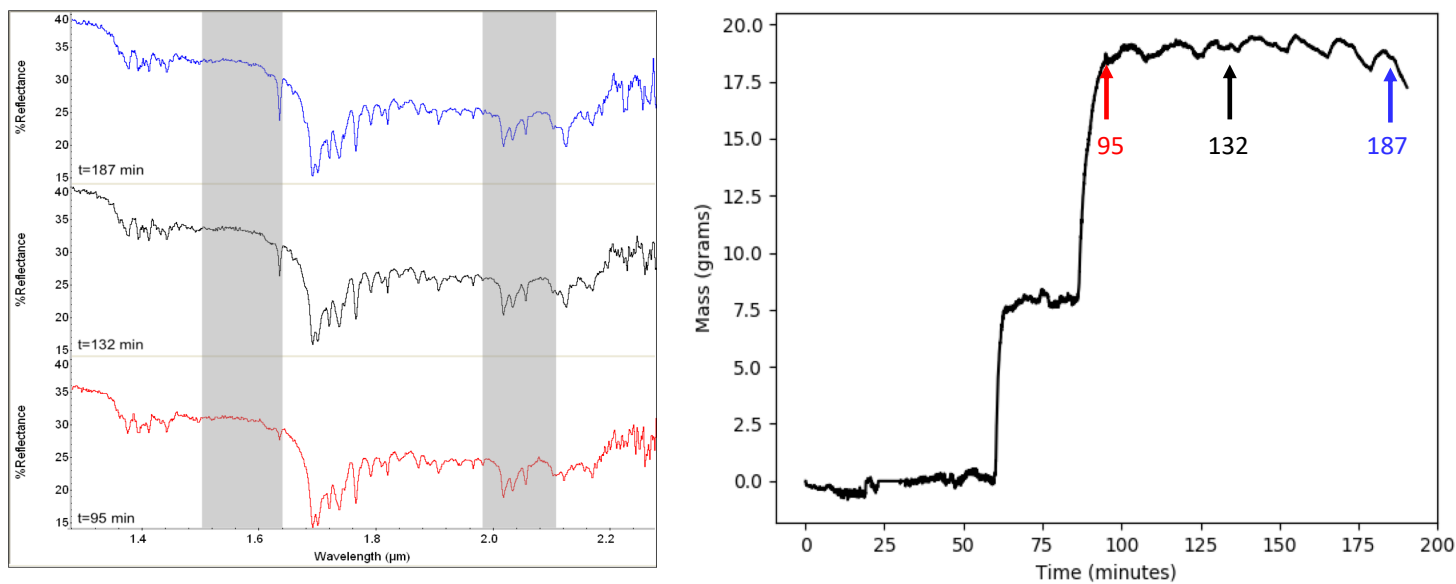

**Figure 7:** Ethane/ethylene spectra, offset for clarity (left). Initial spectrum is shown in red, intermediate in black, and final in blue. Vertical, gray bars indicate VIMS windows. Notice the increase in reflectance throughout the experiment. Mass data from the ethane/ethylene 90 K experiment (right) shows two separate pours due to nonideal chamber conditions in the first pour: one at ~60 minutes and one at ~80 minutes. Spectral analysis only includes data after the second pour. Notice the relatively stable mass from ~100 minutes to the end of the experiment. Colored arrows correspond with the time each spectrum was taken.

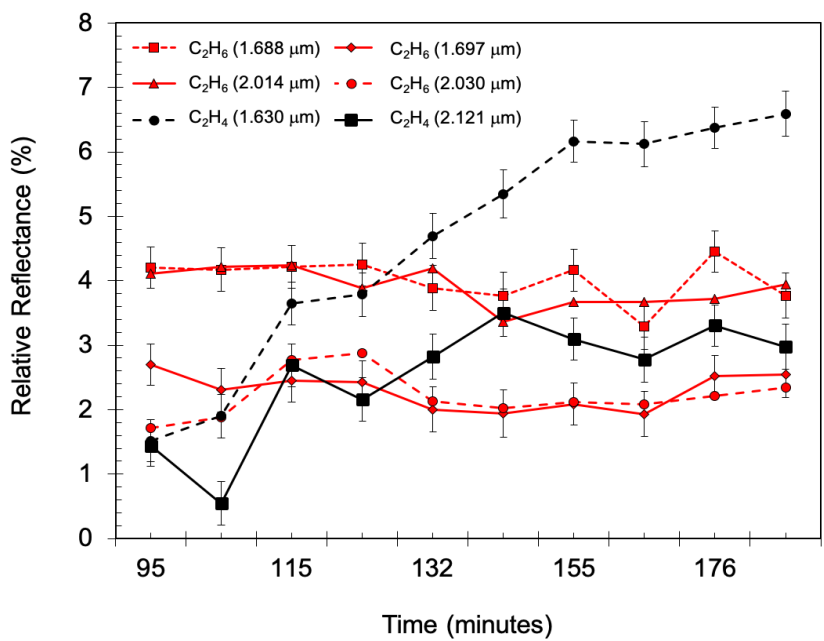

**Figure 8.** Ethane band depths (red) from the 90 K ethane/ethylene experiment show no overall change in characteristic ethane bands (1.688 μm, 1.697 μm, 2.014 μm, and 2.030 μm), while ethylene band depths (black) show an increase in characteristic ethylene bands (1.630 μm and 2.121 μm).

**Forced Evaporation Experiment.** We also performed a second ethane/ethylene experiment, but instead of maintaining Titan temperatures throughout the experiment, the TCB was warmed from ~92 K to ~139 K over a period of 94 minutes to induce evaporation of ethane. See Figure S8 for a temperature profile of the sample over the duration of this experiment.

Figure 9 shows the offset FTIR spectra for this experiment (initial spectrum in red, final spectrum in blue). The spectral results of this experiment show relatively constant band depths for the first ~120 minutes (average temperature of 104 K), then a sudden decrease in both species' band depths starting at 186 minutes (118 K) and ending at 219 minutes (139 K) (Fig. 10). Band depth measurements (Fig. 10) support these spectral results. No red shifts of bands were observed during this experiment.

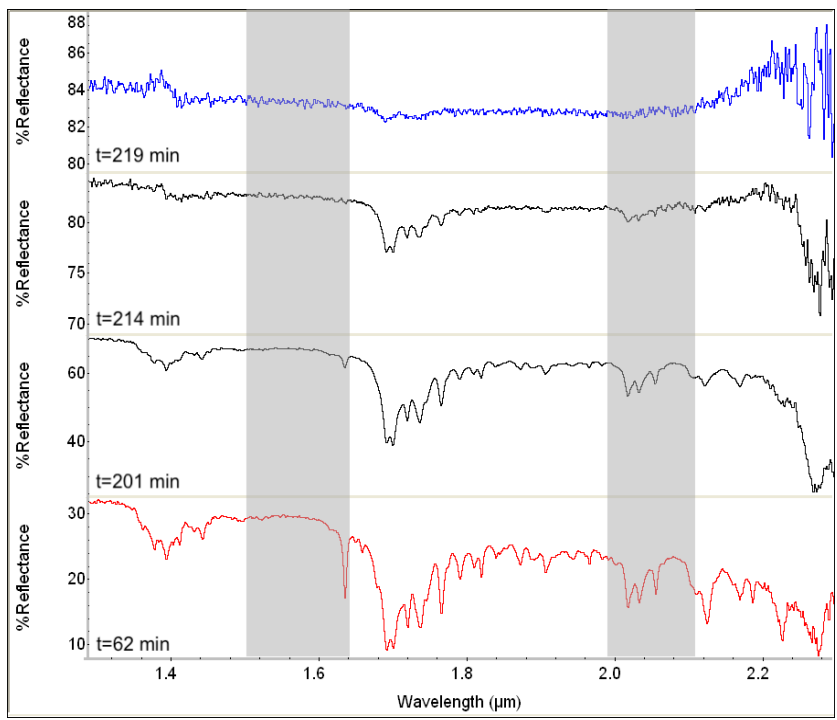

**Figure 9.** Ethane/ethylene spectra, offset for clarity. Initial spectrum is shown in red, intermediate in black, and final in blue. Vertical, gray bars indicate VIMS windows. Notice the increase in reflectance throughout the experiment.

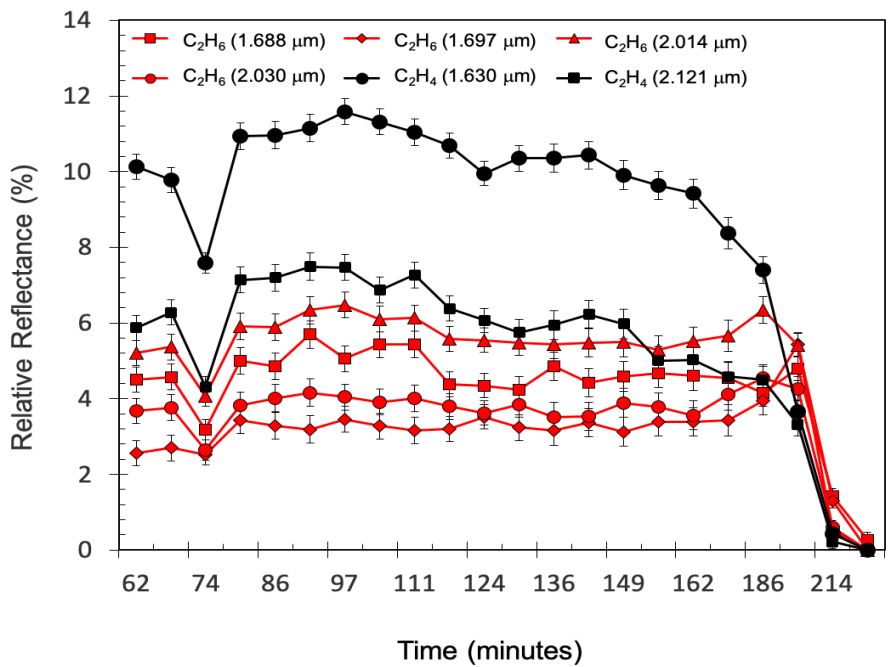

**Figure 10.** Ethane band depths (red) from the force evaporated experiment show constant band depths until force evaporation (~200 minutes), but ethylene band depths (black) began decreasing sooner, around 150 minutes.

**Methane, Ethane, and Ethylene.** Protocol for this experiment included: 50.1% methane, 44.9% ethane, and 5.0% ethylene, initially. Offset reflectance FTIR spectra is shown in Figure 11 (left), with the first sample shown by the red spectrum, and final spectrum shown in blue. The spectra in Figure 11 show variation from the initial spectrum to the final spectrum, which was 0.2% methane, 91.9% ethane, and 7.9% ethylene. The methane in the sample mixture evaporated in this experiment, which explains the percentage difference from initial to final spectrum.

Figure 11 (right) shows the mass data for the methane/ethane/ethylene experiment. The methane/ethane/ethylene sample was poured into the petri dish at ~65 minutes, reaching a maximum mass of ~16 g. An immediate decrease in mass (70-100 minutes) to ~12 g was followed by a stable period (100-150 minutes).

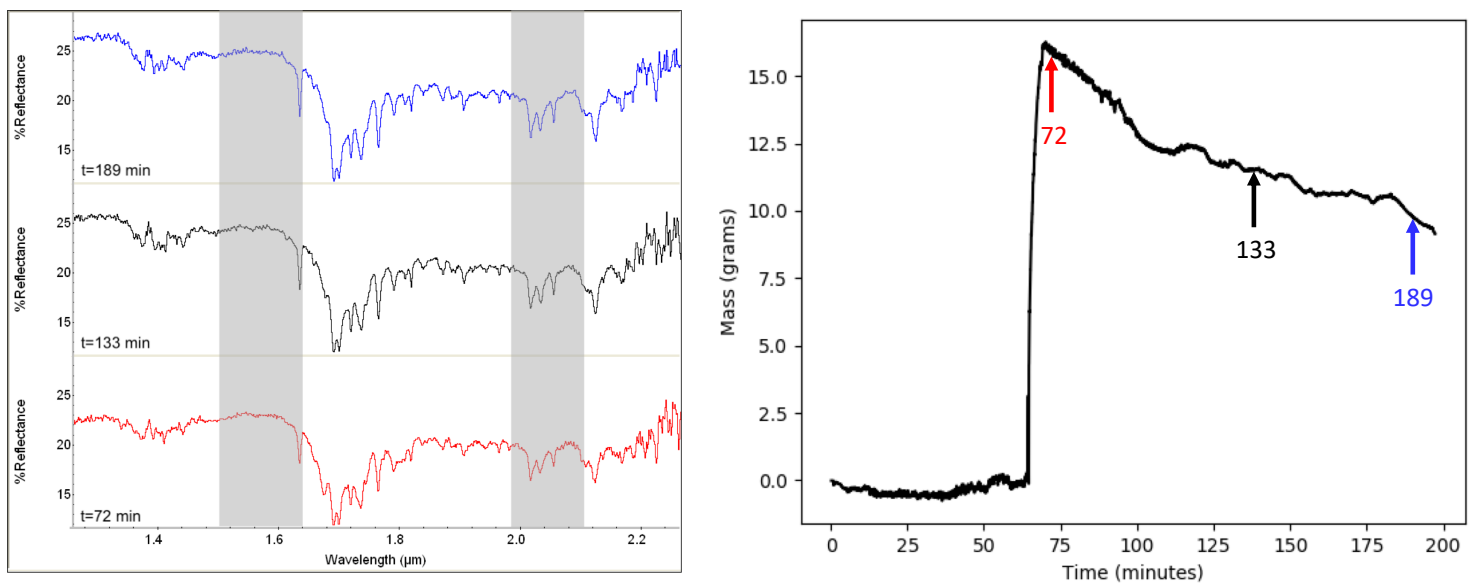

**Figure 11.** Offset reflectance spectra from the methane/ethane/ethylene experiment (left). The initial spectrum is shown in red, intermediate in black, and final in blue. Notice the disappearance of the 1.669 μm methane band throughout the experiment. Mass data from the methane/ethane/ethylene experiment (right) showing that the sample was poured at ~70 minutes, followed by steady state methane evaporation and a semi-stable period of the ethane/ethylene mixture. Colored arrows correspond with the time each spectrum was taken.

Baseline reflectance slightly increased throughout the experiment, beginning at ~10% and ending at ~25%. The methane band depth (1.669 μm) decreased from 0.6% to 0.0% (Fig. 12). Ethane band depths (1.688 μm, 1.697 μm, 2.014 μm, and 2.030 μm) were almost constant at 2.7%, 1.9%, 2.9%, and 1.8% respectively (Fig. 12). Ethylene (1.630 μm and 2.121 μm) band depths slightly increased in relative reflectance from 3.4% to 5.3% and 26% to 3.3%, respectively (Fig. 12). No red shifts of bands were observed in this experiment.

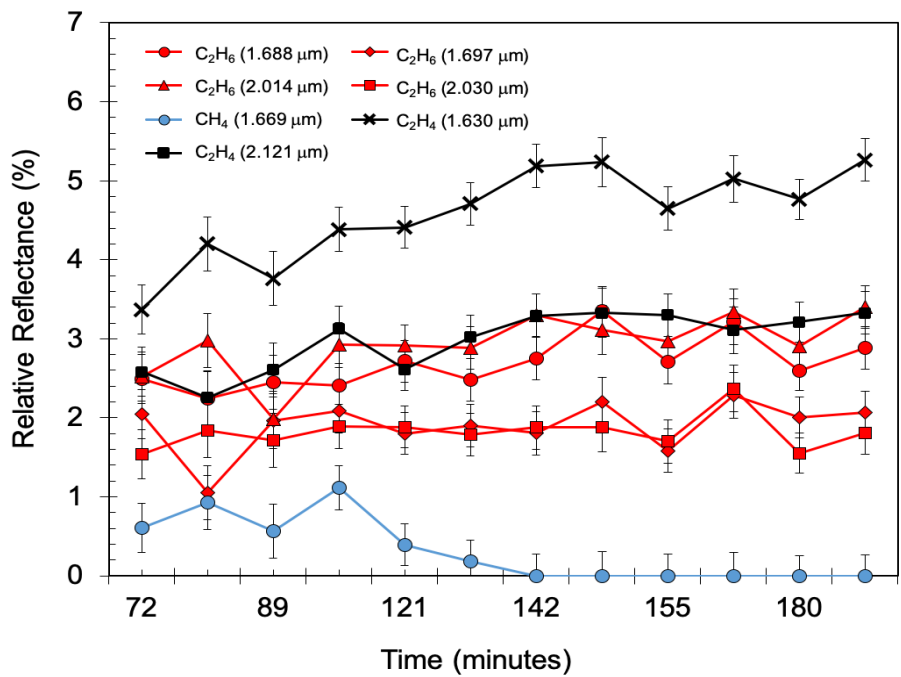

**Figure 12.** Methane band depths (blue) from the methane/ethane/ethylene experiment show a decrease in the methane band (1.669 μm). Ethane band depths (red) show no significant change in ethane bands (1.688 μm, 1.697 μm, 2.014 μm, and 2.030 μm). Ethylene band depths (black) show an increase in both ethylene bands (1.630 μm and 2.121 μm).

**Table 2.** Experimental parameters and evaporation rates. All experiments performed at 1.5 bar.

| Compounds in Experiment | Initial Mole Fraction (%) | Avg. Sample Temp During Spectral Acquisition (K) | Final Mole Fraction (%) | Methane Evaporation Rate (Earth) ($kgm^{-2}s^{-1}$) | Methane Evaporation Rate (Titan) ($kgm^{-2}s^{-1}$) | Ethane Evaporation Rate (Earth) ($kgm^{-2}s^{-1}$) | Ethane Evaporation Rate (Titan) ($kgm^{-2}s^{-1}$) |
|---|---|---|---|---|---|---|---|
| $CH_4$:$C_2H_4$ | 80.5:19.5 | 93.66 | 40.9:59.1 * | $(3.0\pm0.5)\times10^{-4}$ | $(1.1\pm0.2)\times10^{-4}$ | | |
| $CH_4$:$C_2H_4$ | 76.3:23.7 | 90.43 | 41.0:59.0 * | $(2.6\pm0.4)\times10^{-4}$ | $(1.0\pm0.2)\times10^{-4}$ | | |
| $C_2H_6$:$C_2H_4$ | 87.9:12.1 | 93.71 | 76.0:24.0 | | | $<4.9\times10^{-6}$ | $<1.8\times10^{-6}$ |
| $C_2H_6$:$C_2H_4$ | 91.8:8.2 | 93.58 | 84.6:15.4 | | | $<3.5\times10^{-6}$ | $<1.3\times10^{-6}$ |
| $C_2H_6$:$C_2H_4$ & | 65.1:34.9 | 109.48 | & | | | & | & |
| $CH_4$:$C_2H_6$:$C_2H_4$ | 18.1:64.7:17.2 | 95.39 | 8.1:76.4:15.5 | $(3.4\pm0.7)\times10^{-4}$ | $(1.3\pm0.3)\times10^{-4}$ | $<6.8\times10^{-6}$ | $<2.5\times10^{-6}$ |
| $CH_4$:$C_2H_6$:$C_2H_4$ | 50.1:44.9:5.0 | 96.02 | 0.2:91.9:7.9 | $(2.7\pm0.6)\times10^{-4}$ | $(1.0\pm0.2)\times10^{-4}$ | $<7.0\times10^{-6}$ | $<2.6\times10^{-6}$ |
| $CH_4$:$C_2H_6$:$C_2H_4$ | 21.8:74.7:3.5 | 94.37 | 0.2:93.5:6.3 | $(2.5\pm0.9)\times10^{-4}$ | $(0.9\pm0.3)\times10^{-4}$ | $<5.2\times10^{-6}$ | $<1.9\times10^{-6}$ |

* Second number represents saturation value of ethylene.
& Forced evaporation experiment: evaporation rate not relevant.

DISCUSSION

**Evolution of Spectra.** Results for the methane/ethylene experiments are as follows: methane being the more volatile solvent[20] evaporated out of the system, leaving behind a solid residue of ethylene ice. Immediately after this mixture was poured onto the sample dish, the spectrum resembled a mixture of methane and ethylene in liquid phase. However, as additional spectral samples were taken, this liquid methane began evaporating, resulting in precipitation of solid ethylene (Fig. 5) until the only species left in the sample dish was ethylene. The band depths in Figure 5 show that methane values decreased from beginning of the experiment, while ethylene did not change until ~150 minutes. These results, along with the increase in ethylene band depths after 150 minutes (Fig. 5) indicate ethylene precipitation in conjunction with complete methane evaporation. Increasing reflectance and band depth for ethylene could also indicate ethylene slowly becoming the dominant species in the mixture. The overall increase of reflectance and red shifts, which occurred when dissolved ethylene changed to precipitated (solid) ethylene, suggests that the sample becomes enriched in the solid phase, which stems from the solvent's evaporation. The band at 1.666 µm (Fig. 6C) is associated with pure methane. Initial reports of the IR absorption bands of methane describe a band at 1.666 µm as being the partially resolved first harmonic of the 3.33 µm band,[25,26] the first overtone of the $\nu_3$ fundamental,[27] and was classified as $2\nu_3$.[28,29] However, there are discrepancies as to whether these early studies may be referring to the band at 1.669 µm, which is the most apparent methane band in this study and has also been classified as $2\nu_3$. In either case, our spectral results show the persistence of this 1.666 µm methane band throughout the methane/ethylene experiment. This could mean that a small amount of methane may remain in contact with the ethylene evaporite.

In the 90 K ethane/ethylene experiment, there was no significant evaporation of ethane or ethylene, as shown in the spectra and band depth graphs (Figs. 7, 8). This is due to ethane's low vapor pressure, which prevented it from evaporating at Titan surface conditions in our chamber.[20] Stable ethane band depths confirm this interpretation. Most differences between the methane and ethane experiments stem from each solvent's individual vapor pressure. A more significant change is seen in the ethylene band depths, as they increased throughout the experiment. Thus, ethylene forms evaporites slowly with ethane since ethane evaporates slowly. This is contrary to methane, which evaporates quickly, allowing the ethylene evaporite to form much quicker.

In the methane/ethane/ethylene experiment, no ethane evaporation occurred, for the same reason as the ethane/ethylene experiment at Titan temperatures. Given the increased and relatively constant band depths of ethane, we assumed that an insignificant amount of ethane evaporated, which is consistent with our calculated upper limit of ethane evaporation (Table 2). Similar to the other experiments, ethylene band depths increased. The 1.669 μm methane band depth decreased to zero, which indicates complete evaporation from the sample, and no apparent precipitation of ethylene.

**Band Shift and Phase Change.** One objective of studying a single compound during the evaporation process is to identify the transition between dissolved ethylene and solid ethylene solute/solvent interaction. The red shifts in this study are most clearly defined in the ethylene bands at 1.630 μm and 2.121 μm, however we observe the band shifts throughout the ethylene spectrum. These red shifts are indicative of a transition from dissolved ethylene to solid ethylene, and can be explained by the increased density, stronger interactions and C—H bonds, and closer packing in the solid phase when compared to the liquid phase.[30] Initially, ethylene was dissolved in the methane mixture, but as methane evaporated, ethylene reached saturation. At that point, solid ethylene particles were precipitated in the sample dish to form a residual solid, indicated by the red shifts. This transition from dissolved to solid ethylene was not observed in the ethane/ethylene experiment at Titan temperatures because no significant amount of ethane evaporated. Therefore, ethylene remained in a binary mixture with liquid ethane so long as ethane did not completely evaporate. Further, we detected no transition in the methane/ethane/ethylene experiment either. Although methane evaporated out of the system, ethane was still available as a solvent in which ethylene could remain dissolved. This interpretation is also consistent with the fact that no residual solid formed for this experiment, (i.e. lack of band shifts). Given that we measured an average upper limit of ethane evaporation of $<5.5 \times 10^{-6}$ kgm$^{-2}$s$^{-1}$, and modeled ethane evaporation rates on Titan are on the order of $10^{-8}$ kgm$^{-2}$s$^{-1}$,[31] given a long enough period of time, ethane may eventually evaporate completely and leave behind an ethylene evaporite. Alternatively, ethylene may evaporate along with the ethane solvent, as shown in our forced evaporation experiment.

**Solubility.** The initial mole fractions for each experiment were calculated using the analysis outlined in the Spectral Unmixing Model section. Given that the solubility is 56% mole fraction for ethylene in methane and 48% mole fraction for ethylene in ethane,[17] we determined that each of our experiments were well within these solubility limits. The average mole fractions were calculated as 59.1%, 19.7%, and 9.9% in methane, ethane, and a methane/ethane mixtures, respectively. We note that our ethane value is significantly lower than 48% reported in Singh et al.,[17] which is in part due to different experimental protocols between solubility experiments and evaporite experiments. Combining these mole fraction values with the fact that large lakes on Titan may not be saturated in ethylene,[17] we are confident that the ethylene concentrations in our experiments are appropriate for Titan. However, we note that these experimental conditions may not be truly representative, because the composition of Titan's evaporites is still unknown. In the methane/ethylene experiment, we calculated the mole fraction of ethylene during methane evaporation. Figure 5 shows the final methane evaporation occurring between t = 146 and 185 minutes, since this is when the methane band depths decreased to 0% relative reflectance. During this process (t = 165 minutes), the mole fraction of ethylene was measured at 59.0% mole fraction. When compared to 56% mole fraction,[17] our value of 59.0% mole fraction indicates that the solution may be slightly supersaturated.

**Implications for Titan's Lakes.** On Titan, ethylene evaporite deposits would form quicker in lakes or seas dominated by methane. Titan's northern lakes have been predicted to be dominated by methane during the current season,[20,32–35] therefore we may expect the formation of ethylene evaporites in north polar lakes after solvent evaporation. From this, and given the methane and ethane evaporation rates reported in this paper (Table 2), we calculated the time of evaporation and thickness of ethylene evaporite, assuming ethylene dissolution in these lakes. To determine the thickness of an ethylene evaporite layer deposited in one of Titan's northern lakes (e.g. Winnipeg Lacus), we assume lake surface area, depth, and methane-dominated lake composition from Mastrogiuseppe et al.,[35] methane evaporation rates from Table 2, and a methane density of 452 kgm$^{-3}$.[36] For Winnipeg Lacus, it would take ~5 years to produce an ethylene evaporite with a gross total thickness of ~9 m. In the case of a methane/ethylene lake presented here, we deem an evaporite thickness of ~9 m to be reasonable (albeit high) owing to ethylene's high solubility in methane[17] (See details in Supporting Information S12). Additionally, this calculated thickness does not take into account erosion by wind or sublimation processes that may occur, and assumes that the lake is saturated in ethylene, thus at the end of the ~5 year timescale the evaporite would ultimately be less than 9 m thick.

Ethane may also be volatile on timescales longer than a season.[37,38] Ethylene's approximately equal solubility in both methane and ethane suggests evaporite formation in methane-dominated lakes on a seasonal timescale, but possible evaporite formation in ethane-dominated lakes may occur on longer-than-seasonal timescales (i.e. Croll-Milankovich timescale).

The VIMS 5 μm window allowed for observations of spectral units that represent water-ice-poor material on Titan that are most likely evaporites.[2–5] According to Clark et al.,[39] ethylene (92 K) has a band center at 4.90 μm, which falls in the VIMS 5 μm window range. Future experimental studies should extend spectral measurements to 5 μm to determine and characterize absorptions that occur at longer wavelengths.

CONCLUSIONS

By using a unique Titan chamber, we simulated ethylene evaporites in a Titan-like environment. Three types of experiments were performed: methane/ethylene, ethane/ethylene, and methane/ethane/ethylene. Through analysis of spectra, band depths, and temperature, we report that a residual evaporite deposit only formed in the methane/ethylene experiment, due to near complete evaporation of methane. Ethane evaporates much slower than methane at Titan surface conditions, which holds true for our experiments, as no ethylene evaporite formed on the duration of our ethane-dominated experiments. However, these results do not preclude ethylene evaporites from forming in an ethane-dominated lake or sea on Titan, given longer timescales. Our results imply that ethylene may be a good candidate for evaporite deposits at Titan conditions, because ethylene is approximately equally soluble in both methane and ethane solutions. Ethylene evaporites may be restricted to methane-dominated lakes, however, and may not be easily detectable in the 1.6 μm and 2.0 μm VIMS windows.

The recently selected New Frontiers mission, Dragonfly, may be able to constrain the composition of equatorially located Tui and Hotei Regiones, proposed paleo seas that contain the largest 5-μm-bright evaporitic features on the satellite.[4] Dragonfly will land in Titan's dune fields,[40,41] which likely contain material sourced from ancient evaporite deposits. If Tui and Hotei Regiones are indeed evaporitic, as previously implied,[2,4,9] our work here could provide initial constraints on the genesis and processing of these proposed evaporites (i.e. thickness, stratification), in addition to understanding the evolution the dune material underwent before becoming incorporated into the dunes that will be sampled by Dragonfly.

## ASSOCIATED CONTENT

**Supporting Information**

The supporting information is available free of charge. Description of mass analysis code. Spectra, band depths, and mass data from additional experiments.

## AUTHOR INFORMATION

**Corresponding Author**

* Email: ecczapli@email.uark.edu

**ORCID**

Ellen C. Czaplinski: 0000-0002-2046-1416

## ACKNOWLEDGEMENT

The authors greatly acknowledge Shannon MacKenzie for valuable comments in the review process, Marco Mastrogiuseppe for assistance with calculating lake parameters, Walter Graupner for help in the lab, Dustin Laxton for aiding with initial experiments, and Sandeep Singh for insight on experimental protocol. This work was funded by the NASA Earth and Space Science Fellowship (NESSF) Grant # 80NSSC17K0603.

REFERENCES

(1)     Hayes, A.; Aharonson, O.; Callahan, P.; Elachi, C.; Gim, Y.; Kirk, R.; Lewis, K.; Lopes, R.;
        Lorenz, R.; Lunine, J.; et al. Hydrocarbon Lakes on Titan: Distribution and Interaction with a
        Porous Regolith. *Geophys. Res. Lett.* **2008**, *35* (9), 1–6. https://doi.org/10.1029/2008GL033409.

(2)     Barnes, J. W.; Bow, J.; Schwartz, J.; Brown, R. H.; Soderblom, J. M.; Hayes, A. G.; Vixie, G.; Le
        Mouélic, S.; Rodriguez, S.; Sotin, C.; et al. Organic Sedimentary Deposits in Titan's Dry
        Lakebeds: Probable Evaporite. *Icarus* **2011**, *216* (1), 136–140.
        https://doi.org/10.1016/j.icarus.2011.08.022.

(3)     Barnes, J. W. A 5-Micron-Bright Spot on Titan: Evidence for Surface Diversity. *Science.* **2005**,
        *310* (5745), 92–95. https://doi.org/10.1126/science.1117075.

(4)     MacKenzie, S.; and Barnes, J. Compositional Similarities and Distinctions between Titan's
        Evaporitic Terrains. *Astrophys. J.* **2016**, *821* (17). https://doi.org/10.3847/0004-637X/821/1/17.

(5)     Barnes, J. W.; Brown, R. H.; Soderblom, J. M.; Soderblom, L. A.; Jaumann, R.; Jackson, B.; Le
        Mouélic, S.; Sotin, C.; Buratti, B. J.; Pitman, K. M.; et al. Shoreline Features of Titan's Ontario
        Lacus from Cassini/VIMS Observations. *Icarus* **2009**, *201* (1), 217–225.
        https://doi.org/10.1016/j.icarus.2008.12.028.

(6)     Cornet, T.; Bourgeois, O.; Le Mouélic, S.; Rodriguez, S.; Lopez Gonzalez, T.; Sotin, C.; Tobie,
        G.; Fleurant, C.; Barnes, J. W.; Brown, R. H.; et al. Geomorphological Significance of Ontario
        Lacus on Titan: Integrated Interpretation of Cassini VIMS, ISS and RADAR Data and
        Comparison with the Etosha Pan (Namibia). *Icarus* **2012**, *218* (2), 788–806.
        https://doi.org/10.1016/j.icarus.2012.01.013.

(7)     Cordier, D.; Barnes, J. W.; Ferreira, A. G. On the Chemical Composition of Titan's Dry Lakebed
        Evaporites. *Icarus* **2013**, *226* (2), 1431–143. https://doi.org/10.1016/j.icarus.2013.07.026.

(8)     Cordier, D.; Cornet, T.; Barnes, J. W.; MacKenzie, S. M.; Le Bahers, T.; Nna-Mvondo, D.;
        Rannou, P.; Ferreira, A. G. Structure of Titan's Evaporites. *Icarus* **2016**, *270*, 41–56.
        https://doi.org/10.1016/j.icarus.2015.12.034.

(9)     MacKenzie, S. M.; Barnes, J. W.; Sotin, C.; Soderblom, J. M.; Le Mouélic, S.; Rodriguez, S.;
        Baines, K. H.; Buratti, B. J.; Clark, R. N.; Nicholson, P. D.; et al. Evidence of Titan's Climate
        History from Evaporite Distribution. *Icarus* **2014**, *243*, 191–207.
        https://doi.org/10.1016/j.icarus.2014.08.022.

(10)    Cable, M. L.; Vu, T. H.; Hodyss, R.; Choukroun, M.; Malaska, M. J.; Beauchamp, P.
        Experimental Determination of the Kinetics of Formation of the Benzene-Ethane Co-Crystal and
        Implications for Titan. *Geophys. Res. Lett.* **2014**, *41* (15), 5396–5401.
        https://doi.org/10.1002/2014GL060531.

(11)    Cable, M. L.; Vu, T. H.; Maynard-Casely, H. E.; Choukroun, M.; Hodyss, R. The Acetylene-
        Ammonia Co-Crystal on Titan. *ACS Earth Sp. Chem.* **2018**, *2*(4), 366-375.
        https://doi.org/10.1021/acsearthspacechem.7b00135.

(12)    Czaplinski, E.; Farnsworth, K.; Laxton, D.; Chevrier, V.; Heslar, M.; Singh, S. Experimental
        Results of Evaporite Deposits on Titan Using a Surface Simulation Chamber. *LPSC XLVIII* **2017**,
        *2017*, 1537.

(13)    Czaplinski, E.; Farnsworth, K.; Gilbertson, W.; Chevrier, V. Experimental Studies of Ethylene
        and Benzene Evaporites on Titan. *LPSC XLIX* **2018**, *2018*, 1480.

(14)    Malaska, M. J.; Hodyss, R. Dissolution of Benzene, Naphthalene, and Biphenyl in a Simulated
        Titan Lake. *Icarus* **2014**, *242*, 74–81. https://doi.org/10.1016/j.icarus.2014.07.022.

(15)    Vu, T. H.; Cable, M. L.; Choukroun, M.; Hodyss, R.; Beauchamp, P. Formation of a New
        Benzene-Ethane Co-Crystalline Structure under Cryogenic Conditions. *J. Phys. Chem.* **2014**, *118*

(23), 4087–4094. https://doi.org/10.1021/jp501698j.

(16)    Roe, H. G.; de Pater, I.; McKay, C. P. Seasonal Variation of Titan's Stratospheric Ethylene
        ($C_2H_4$) Observed. *Icarus* **2004**, *169* (2), 440–461. https://doi.org/10.1016/j.icarus.2004.01.002.

(17)    Singh, S.; Combe, J.-P.; Cordier, D.; Wagner, A.; Chevrier, V. F.; McMahon, Z. Experimental
        Determination of Acetylene and Ethylene Solubility in Liquid Methane and Ethane: Implications
        to Titan's Surface. *Geochim. Cosmochim. Acta* **2017**, *208*, 86–101.
        https://doi.org/10.1016/j.gca.2017.03.007.

(18)    Wasiak, F. C.; Luspay-Kuti, A.; Welivitiya, W. D. D. P.; Roe, L. A.; Chevrier, V. F.; Blackburn,
        D. G.; Cornet, T. A Facility for Simulating Titan's Environment. *Adv. Sp. Res.* **2013**, *51* (7),
        1213–1220. https://doi.org/10.1016/j.asr.2012.10.020.

(19)    Luspay-Kuti, A.; Chevrier, V. F.; Wasiak, F. C.; Roe, L. A.; Welivitiya, W. D. D. P.; Cornet, T.;
        Singh, S.; Rivera-Valentin, E. G. Experimental Simulations of $CH_4$ Evaporation on Titan.
        *Geophys. Res. Lett.* **2012**, *39* (23), 2–6. https://doi.org/10.1029/2012GL054003.

(20)    Luspay-Kuti, A.; Chevrier, V. F.; Cordier, D.; Rivera-Valentin, E. G.; Singh, S.; Wagner, A.;
        Wasiak, F. C. Experimental Constraints on the Composition and Dynamics of Titan's Polar
        Lakes. *Earth Planet. Sci. Lett.* **2015**, *410*, 75–83. https://doi.org/10.1016/j.epsl.2014.11.023.

(21)    Blunt, V. M.; Cedeño, D. L.; Manzanares I, C. Vibrational Overtone Spectroscopy of Methane in
        Liquid Argon Solutions. *Mol. Phys.* **1997**, *91*, 3–17. https://doi.org/10.1080/002689797171698.

(22)    Hepp, M.; Herman, M. Vibration-Rotation Bands in Ethane. *Mol. Phys.* **2009**, *98* (2000), 57–61.
        https://doi.org/10.1080/00268970009483269.

(23)    Brock, A.; Mina-camilde, N.; I, C. M. Vibrational Spectroscopy of C-H Bonds of $C_2H_4$ Liquid
        and $C_2H_4$ in Liquid Argon Solutions. *J. Phys. Chem.* **1994**, *98*, 4800–4808.

(24)    Mitri, G.; Showman, A. P.; Lunine, J. I.; Lorenz, R. D. Hydrocarbon Lakes on Titan. *Icarus* **2007**,

*186* (2), 385–394. https://doi.org/10.1016/j.icarus.2006.09.004.

(25) Cooley, J. P. The Infra-Red Absorption Bands of Methane. *Astrophys. J.* **1925**, *62*, 73–83.

(26) Ellis, J. W. New Infra-Red Absorption Bands of Methane. *Physics (College. Park. Md).* **1927**, *13*, 202–207.

(27) Nelson, B. R. C.; Plylerl, E. K.; Benedict, W. S. Absorption Spectra of Methane in the Near Infrared. **1948**, *41*, 615–621.

(28) Moorhead, J. G. THE NEAR INFRARED ABSORPTION SPECTRUM OF METHANE. *Phys. Rev.* **1932**, *39*, 83–88.

(29) Norris, W. V; Unger, H. J. Infrared Absorption Bands of Methane. *Pysical Rev.* **1933**, *43*, 467–472.

(30) Abramczyk, H.; Paradowska-Moszkowska, K. The Correlation between the Phase Transitions and Vibrational Properties by Raman Spectroscopy : Liquid-Solid β and Solid β -Solid α Acetonitrile Transitions. *Chem. Phys.* **2001**, *265*, 177–191. https://doi.org/https://doi.org/10.1016/S0301-0104(01)00271-3.

(31) Tokano, T. Limnological Structure of Titan's Hydrocarbon Lakes and Its Astrobiological Implication. *Astrobiology* **2009**, *9* (2), 147–164. https://doi.org/10.1089/ast.2007.0220.

(32) Mastrogiuseppe, M.; Poggiali, V.; Hayes, A.; Lorenz, R.; Lunine, J.; Picardi, G.; Seu, R.; Flamini, E.; Mitri, G.; Notarnicola, C.; et al. The Bathymetry of a Titan Sea. *Geophys. Res. Lett.* **2014**, *41*, 1432–1437. https://doi.org/10.1002/ 2013GL058618. Received.

(33) Mastrogiuseppe, M.; Hayes, A.; Poggiali, V.; Seu, R.; Lunine, J. I.; Hofgartner, J. D. Radar Sounding Using the Cassini Altimeter: Waveform Modeling and Monte Carlo Approach for Data Inversion of Observations of Titan's Seas. *Int. J. Health Serv.* **2016**, *54*, 5646–5656. https://doi.org/10.2190/RM36-JLXA-6WUY-KDQ2.

(34)  Mastrogiuseppe, M.; Poggiali, V.; Hayes, A. G.; Lunine, J. I.; Seu, R.; Di Achille, G.; Lorenz, R. D. Cassini Radar Observation of Punga Mare and Environs: Bathymetry and Composition. *Earth Planet. Sci. Lett.* **2018**, *496*, 89–95. https://doi.org/10.1016/j.epsl.2018.05.033.

(35)  Mastrogiuseppe, M.; Poggiali, V.; Hayes, A. G.; Lunine, J. I.; Seu, R.; Mitri, G.; Lorenz, R. D. Deep and Methane-Rich Lakes on Titan. *Nat. Astron.* **2019**, *3*, 535-542. https://doi.org/10.1038/s41550-019-0714-2.

(36)  Malaska, M. J.; Hodyss, R.; Lunine, J. I.; Hayes, A. G.; Hofgartner, J. D.; Hollyday, G.; Lorenz, R. D. Laboratory Measurements of Nitrogen Dissolution in Titan Lake Fluids. *Icarus* **2017**, *289*, 94–105. https://doi.org/10.1016/j.icarus.2017.01.033.

(37)  Lunine, J. I.; Stevenson, D. J.; Yung, Y. L. Ethane Ocean on Titan. *Science* **1983**, *222*, 1229–1230.

(38)  Mousis, O.; Lunine, J. I.; Hayes, A. G.; Hofgartner, J. D. The Fate of Ethane in Titan's Hydrocarbon Lakes and Seas. *Icarus* **2016**, *270*, 37–40. https://doi.org/10.1016/j.icarus.2015.06.024.

(39)  Clark, R. N.; Curchin, J. M.; Barnes, J. W.; Jaumann, R.; Soderblom, L.; Cruikshank, D. P.; Brown, R. H.; Rodriguez, S.; Lunine, J.; Stephan, K.; et al. Detection and Mapping of Hydrocarbon Deposits on Titan. **2010**, *115*. https://doi.org/10.1029/2009JE003369.

(40)  Lorenz, R. D.; Turtle, E. P.; Barnes, J. W.; Trainer, M. G.; Adams, D. S.; Hibbard, K. E.; Sheldon, C. Z.; Zacny, K.; Peplowski, P. N.; Lawrence, D. J.; et al. Dragonfly : A Rotorcraft Lander Concept for Scientific Exploration at Titan. **2017**, *4*, 1–14.

(41)  Turtle, E. P.; Barnes, J. W.; Trainer, M. G.; Lorenz, R. D.; Hibbard, K. E.; Adams, D. S.; Bedini, P.; Brinckerhoff, W. B.; Ernst, C.; Freissinet, C.; et al. Dragonfly: In Situ Exploration of Titan's Organic Chemistry and Habitability. *49th LPSC* **2018**, Abstract 1641.

https://doi.org/10.1109/AERO.2013.6497165.

TOC Graphic

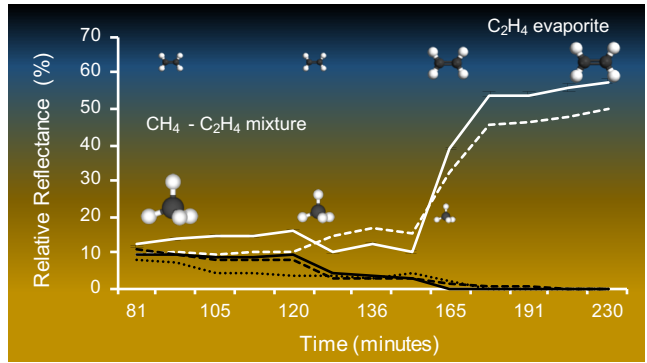